\documentclass{article}
\usepackage{amsmath}
\usepackage{amssymb}
\usepackage{graphicx}
\begin{document}
\begin{center}
\LARGE 
\textbf{Dynamical Correspondence in a Generalized Quantum Theory}
\\[0,5 cm]
\normalsize
Gerd Niestegge
\\[0,3 cm]
\scriptsize
Fraunhofer ESK, Hansastr. 32, 80686 M\"unchen, Germany
\\
gerd.niestegge@esk.fraunhofer.de, 
gerd.niestegge@web.de
\\[0,5 cm]
\end{center}
\normalsize
\normalfont \itshape Abstract. \normalfont
In order to figure out why quantum physics needs the complex Hilbert space, 
many attempts have been made to distinguish the C*-algebras and von Neumann algebras in 
more general classes of abstractly defined Jordan algebras (JB- and JBW-algebras).
One particularly important distinguishing property
was identified by Alfsen and Shultz and is the existence of a
dynamical correspondence. It reproduces the dual role of the selfadjoint operators
as observables and generators of dynamical groups
in quantum mechanics.
In the paper, this concept is extended 
to another class of nonassociative algebras, arising
from recent studies of the quantum logics with a conditional probability
calculus and particularly of those that rule out third-order interference.
The conditional probability calculus is a mathematical model of the 
L\"uders-von Neumann quantum measurement process, and 
third-order interference is a property of the conditional probabilities
which was discovered by R. Sorkin in 1994 and which is ruled out 
by quantum mechanics.
It is shown then that the postulates that a dynamical
correspondence exists and that the square of any algebra element is positive
still characterize, in the class considered, those algebras 
that emerge from the selfadjoint parts of C*-algebras equipped with the Jordan product.
Within this class, 
the two postulates thus result in ordinary quantum mechanics 
using the complex Hilbert space or, vice versa, a genuine generalization of 
quantum theory must omit at least one of them.
\\[0,3 cm]
\textit{Key Words.} Order derivations, positive groups, operator algebras, 
Lie algebras, foundations of quantum mechanics 
\\[0,3 cm]
\textit{PACS.} 03.65Fd, 03.65:Ta, 02.30.Tb
\newline
\textit{MSC.} 46L70, 81P10

\section{Introduction}

In quantum mechanics, the selfadjoint operators play a dual role; they represent
observables  - the measurable physical quantities of the system under consideration -
as well as generators of dynamical groups - describing the time evolution of the system.
In a more general setting, this dual role is reproduced by a
dynamical correspondence allocating a 
generator of a dynamical group on an ordered Banach space
to each element of the space; in this case, the group generators are 
skew order derivations. Dynamical correspondences were 
introduced by Alfsen and Shultz \cite{AS98,AS03} as a mean to
distinguish the selfadjoint parts of the C*-algebras and von Neumann algebras among 
Jordan algebras of a more general type - the JB- and JBW-algebras -
and to figure out why quantum physics needs the complex Hilbert space. 
Dynamical correspondences base upon Connes' notion of order derivations \cite{Con74}. 

In the present paper, which is a sequel of Ref. \cite {Nie14},
the notion of a dynamical correspondence is extended to 
another class of nonassociative algebras 
comprising the JBW algebras and 
arising from recent studies
of the quantum logics with a conditional probability
calculus and particularly of those that rule out 
third-order interference \cite {Nie08, Nie12, Nie14}. 
The conditional probability calculus is a mathematical model of the 
L\"uders-von Neumann quantum measurement process.  
Third-order interference is a property of the conditional probabilities
which was discovered by R. Sorkin in 1994 and which is ruled out 
by quantum mechanics \cite{Sor94}.

Below, it will be shown that the postulates that a dynamical
correspondence exists and that the square of any algebra element is positive
still characterize, in the class considered, those algebras 
that emerge from the selfadjoint parts of C*-algebras equipped with the Jordan product.

The class of nonassociative algebras is defined in section 2 and, how 
operator algebras and Jordan algebras fit into this setting, is explained 
in section 3. Order derivations are considered 
in section 4, before turning to the dynamical correspondences and the main
results in section 5. 

\section{The ordered Banach algebra}

Let $A$ be a complete order-unit space with distinguished order-unit $\mathbb{I}$ \cite{AS03}. 
Its unit interval
$\left[0,\mathbb{I}\right]$ := $\left\{a \in A: 0\leq a \leq \mathbb{I} \right\}$ = 
$\left\{a \in A: 0 \leq a\ \mbox{and}\ \left\| a \right\| \leq 1 \right\}$
is a convex set. As usual,
an element in a convex set is called an extreme point of this set
if it is not any convex combination of two other elements in this set.
The set of extreme points of the unit interval is denoted by $ext\left[0,\mathbb{I}\right]$.
It includes the order-unit $\mathbb{I}$
and forms a quantum logical structure 
with the orthogonality relation
$e \bot f :\Longleftrightarrow e + f \in ext\left[0,\mathbb{I}\right]$ 
and with the orthocomplementation 
$e' := \mathbb{I} - e$ 
($e,f \in ext\left[0,\mathbb{I}\right]$).
Its elements (the extreme points of $\left[0,\mathbb{I}\right]$) are called \textit{events}.

A \textit{state} $\mu$ on this quantum logic
allocates the probability $ \mu(f) \in [0,1]$ to each 
event $f$ and
is an orthogonally additive function from
$ext\left[0,\mathbb{I}\right]$ to the non-negative real numbers with 
$\mu(\mathbb{I}) = 1$.
The \textit{conditional 
probability} of an event $f$ under another event $e$ in the state $\mu$ with $\mu(e) \neq 0$ is the 
updated probability for $f$ after the outcome of
a first measurement has been the event $e$; it is denoted 
by $ \mu(f | e) $. Mathematically, it is defined by the
conditions that the map $ext\left[0,\mathbb{I}\right] \ni f \rightarrow \mu(f | e)$
is a state on $ext\left[0,\mathbb{I}\right]$ and that the identity 
$ \mu(f | e) = \mu(f)/\mu(e)$ holds for all events 
$f \in ext\left[0,\mathbb{I}\right]$ with $f \leq e$. 

Note below that an operator $S : A \rightarrow A$
(or a function $\rho : A \rightarrow \mathbb{R}$)  
is called \textit{positive} if $S(a) \geq 0$ ($\rho(a) \geq 0$)
for all $a \in A$ with $a \geq 0$.

In the remaining part of this paper, it shall be assumed that $A$
is a complete order-unit space with order-unit $\mathbb{I}$ and that 
the following four conditions (A), (B), (C) and (D) are satisfied:
\\[1 cm]
\textbf{(A)} There is a bilinear multiplication $\Box$ on $A$ 
with $\mathbb{I}$ as multiplicative identity,
$\left\|a \Box b\right\| \leq \left\|a\right\|\:\left\|b\right\|$ for $a,b \in A$
and $e^{2} := e \Box e = e $ for all
$e \in ext\left[0,\mathbb{I}\right]$.
\\
\textbf{(B)} Define $T_a b := a \Box b$ for $a,b \in A$ and $U_e := 2 T_e^{2} - T_e$
for $e \in ext\left[0,\mathbb{I}\right]$. Each linear operator $U_e$ is 
positive and its range $U_e A$ is the closed linear hull of 
$\left\{ f \in ext\left[0,\mathbb{I}\right] \ | \ f \leq e \right\}$.
\\
\textbf{(C)}  Every state $\mu$ on the quantum logic $ext\left[0,\mathbb{I}\right]$
has a positive linear extension on $A$ which is again denoted by $\mu$
(Note that this extension is unique 
since the positive linear functionals on an order-unit space are continuous 
and since, by applying (B) to $e = \mathbb{I}$, $A$ is the closed linear hull 
of $ext\left[0,\mathbb{I}\right]$. This also implies that $\left[0,\mathbb{I}\right]$ contains
sufficiently many extreme points). 
\\
\textbf{(D)} If $\mu$ is a state and $0 \leq a \in A$ with $\mu(a)=0$,
then $\mu(a \Box b) = 0$ for all $b \in A$. 
\\[0,3 cm]
Note that, generally, the product $\Box$ is neither commutative nor associative, 
$\mu(b \Box a) = 0$ does not hold in condition (D), and
the square $a^{2} = a \Box a $ of an element $a \in A$ is not positive.
\\[0,3 cm]
Lemma 1: Suppose $e,f \in ext\left[0,\mathbb{I}\right]$ and $a \in A$; then: 
\newline
(i) $U_{e}^{2} = U_e$ and $U_e U_{e'} = U_{e'}U_e = 0$. 
\newline
(ii) $U_e f = f$ and $U_{e'}f = 0$ for $f \leq e$. Moreover, $U_e f = 0$ and $U_{e'}f = f$ for $e \bot f$.
\newline
(iii) $e \Box a = T_e a = (a + U_e a - U_{e'} a)/2$. 
\newline
(iv) $e \Box f = T_e f = f$ and $e' \Box f = T_{e'} f = 0$ for $f \leq e$. 
Moreover, $e \Box f = T_e f = 0$ and $e' \Box f = T_{e'} f = f$ for $e \bot f$.
\\[0,3 cm]
Proof. 
(i) Assume that $\mu$ is a state with $\mu(e) \neq 0$. Then
$\nu (f)  := \mu (U_e f)/\mu(e)$ for the events $f$ defines
a state $\nu$.  The identity $e^{2}=e$ implies $\nu(e)=1$ and thus $\nu(e')=0$. By (D),
$0 = \nu(e' \Box x)= \mu(U_e (e' \Box x)) / \mu (e)$ 
and thus $0 = \mu(U_e (e' \Box x))$ for any $x$ in $A$. 

If $\mu(e)=0$, then $0 = \mu (T_e y)$ 
for all $y \in A$ by (D) and $0 = \mu (U_e y)$ by the definition of $U_e$. 
Thus again $0 = \mu(U_e (e' \Box x)) $ for any $x \in A$. 

Therefore $0 = U_e (e' \Box x)$ for any $x \in A$. 
This means $0 = U_e T_{e'} = U_e - U_e T_e$ and $U_e = U_e T_e$. Using again the 
definition of $U_e$ and $U_{e'}$ finally gives $U^{2}_{e} = U_e (2T^{2}_{e} - T_e) = U_e$
and $U_e U_{e'} = U_e (2T_{e'}^{2} - T_{e'}) = 0$.
Replacing $e$ by $e'$ yields $U_{e'}U_e = 0$.

(ii) Suppose $f \leq e$. Then $U_f A \subseteq U_e A$ by (B).
Therefore $U_e U_f = U_f$ by (i) and
$U_e f = U_e U_f \mathbb{I} = U_f \mathbb{I} = f$.
Moreover, $0 \leq U_{e'} f \leq U_{e'} e = 0$ and hence $U_{e'} f = 0$.
If $e \bot f$, then $f \leq \mathbb{I} - e = e'$ and, 
replacing $e$ by $e'$ in the first part of (ii), yields 
$U_e f = 0$ as well as $U_{e'} f = f$.

(iii) Note that $U_{e'} a = 2 e' \Box (e' \Box a) - e' \Box a 
= 2 (\mathbb{I} - e) \Box (e' \Box a) - e' \Box a 
= 2 (e' \Box a) - 2 e \Box (e' \Box a) - e' \Box a  
= e' \Box a - 2 e \Box (e' \Box a)  
= a - e \Box a - 2 e \Box a + 2 e \Box (e \Box a)
= a - 3 T_e a + 2 T_e^{2} a$ and therefore,
$a + U_e a - U_{e'} a
= a + 2 T_e^{2} a - T_e a -  a + 3 T_e a - 2 T_e^{2} a
= 2 T_e a$.

(iv) follows from (ii) and (iii). $\blacksquare$
\\[0,3 cm]
Lemma 2: With the above assumptions, the conditional probability 
$ext\left[0,\mathbb{I}\right] \ni f \rightarrow \mu(f | e)$
exists and is uniquely determined for any state $\mu$ and event $e$
with $\mu(e) \neq 0$. Moreover, $\mu(f | e) = \mu (U_e f)/\mu(e)$.
\\[0,3 cm]
Proof. Suppose $\mu(e) \neq 0$ and define $\nu_1(f) := \mu (U_e f)/\mu(e)$
for $f \in ext\left[0,\mathbb{I}\right]$. By Lemma 1 (ii), $\nu_1$
is a conditional probability.

Assume that $\nu_2$ is a second conditional probability under $e$ in the state $\mu$.
Then $\nu_2 (e) = 1$ and $\nu_2 (e') = 0$. By (D), $\nu_2 (e' \Box x) = 0$ for all
$x$ in $A$ or, equivalently, $\nu_2 (T_e x) = \nu_2 (e \Box x) = \nu_2 (x)$. This means
that $\nu_2$ is invariant under $T_e$ and therefore under $U_e = 2 T_e^{2} - T_e$.

For any $x \in A$ then 
$\nu_2(x) = \nu_2(U_e x)$. Since $U_e x$ lies in 
the closed linear hull of $\left\{f \in ext\left[0,\mathbb{I}\right] \ | \ f \leq e \right\}$, 
the characteristics of the conditional probability and 
the continuity and linearity of $\nu_2$ imply
$\nu_2(x) = \nu_2(U_e x) = \mu (U_e x)/\mu(e) = \nu_1 (x)$.
Therefore, $\nu_1 = \nu_2$. $\blacksquare$
\\[0,3 cm]
The mathematical structure defined in this section has two important aspects. On the one hand,
the next section will show that it covers the operator algebras used in quantum physics, 
but is more general. 

On the other hand, it stems from recent studies of the quantum logics 
with a conditional probability calculus (i.e. with a reasonable model of the 
L\"uders-von Neumann quantum measurement process) and particularly of those 
that rule out third-order interference \cite{Nie12, Nie14}. 
It can thus be regarded as a generalized quantum theory. 
However, note that
it does not encompass the most general case studied in Refs. \cite{Nie12, Nie14}. 
For instance, generally,  
the quantum logic need not coincide with 
the extreme points of the unit interval, and condition (D) is not satisfied 
for all positive elements $a$ in $A$, but only if $a$ lies in the quantum logic.
Moreover, in the infinite-dimensional case, the norm topology and certain weak 
topologies must be distinguished and the norm topology must be replaced by a weak topology 
is some cases to cover the most general situation.

\section{Operator algebras}

The formally real Jordan algebras were introduced in 1934 by Jordan, von Neumann and Wigner \cite{JNW34}. 
Forty years later, this theory was extended by Alfsen, Shultz, St\o rmer and others to include infinite dimensional algebras;
these are the so-called JB-algebras and JBW-algebras. 
The monograph \cite{AS03} contains a comprehensive representation of the theory of the JB-/JBW-algebras. 
The selfadjoint part of
a C*-algebra, equipped with the Jordan product $a \circ b = (ab + ba)/2$, is a JB-algebra,
and the selfadjoint part of a von Neumann algebra is a JBW-algebra.
All these algebras are order-unit spaces.

A JBW-algebra $A$ without type $I_2$ part satisfies the four conditions
in section 2, with the Jordan product $\circ$ playing the role of the $\Box$-product. 
Condition (A) is the well-known fact that the extreme points of the positive part
of the unit ball in a JBW-algebra coincide with the idempotent elements of the JBW-algebra \cite{AS03}. 
The positivity of the $U_e$ is a rather non-trivial result in the theory of the 
JBW-algebras \cite{AS03} and, together with the spectral theorem, it implies 
the remaining part of condition (B).
Condition (C) is the extension of Gleason's theorem
to JBW-algebras which holds iff $A$ does not contain a type $I_2$ part \cite{BW89}.
Condition (D) follows from the Cauchy-Schwarz inequality for the positive
linear functionals and the fact that, by the spectral theorem, 
$\mu(a)=0$ implies $\mu(a^{2})=0$ for $0 \leq a$. A more explicit elaboration of these
considerations can be found in \cite{Nie01}.

In the specific situation where $A$ is the selfadjoint part
of a von Neumann algebra $M$, the operators $U_e$ get the familiar shape
$U_e a = eae$ ($a,e \in A$ and $e^{2}=e$),
which also reveals the link to the L\"uders-von Neumann quantum measurement process.
In this case, the positivity of $U_e$ follows from $U_e a = eae = (a^{1/2}e)^{*}(a^{1/2}e) \geq 0$
for $a \geq 0$.

\section{Order derivations}

A bounded linear operator $D:A \rightarrow A$ is called 
an \textit{order derivation} 
if $e^{tD}$ is positive for any real number $t$. 
The order derivations are generators of positive groups.
They were introduced by Connes \cite{Con74}.
Most interesting are those positive groups, 
which leave the order-unit invariant for all $t$, since
they entail automorphism groups of the state space \cite{Nie14}; this holds 
when the generator $D$ satisfies the condition $D(\mathbb{I})=0$.
Such a derivation $D$ describes the dynamical evolution satisfying the simple
linear differential equation $ \frac{d}{dt}x_t = Dx_t$ $(x_t \in A)$.
Any physical theory with a reversible time evolution
should include one-parameter automorphism groups and therefore at least
some derivations $D$ with $D(\mathbb{I})=0$. 
Generally, they need not be bounded, but note that only bounded derivations
are considered in this paper. 

The following lemma provides a very useful general characterization of   
order derivations; it was first used by Connes in a more specific context \cite{Con74}
and then generalized by Evans and Hanche-Olsen \cite{Evans79} (see also \cite{AS01}).
\\[0,3 cm]
Lemma 3. Let $D$ be a bounded linear operator 
from $A$ into $A$. Then the following are equivalent:
\newline
(i) $D$ is an order derivation.
\newline
(ii) If $\mu$ is a state and $0 \leq x \in A$ with $\mu (x) = 0$, then $\mu (Dx) = 0$. 
\\[0,3 cm]
By Lemma 3 and condition (D) in section 2, the 
right-hand side multiplication operators $R_a$ 
with $R_a x := x \Box a$, $x \in A$, are order derivations for all $a \in A$.
They are called \textit{selfadjoint}. Of course, $R_a(\mathbb{I}) = a$. 
The more interesting order derivations $D$ with $D(\mathbb{I}) = 0$
are called \textit{skew}. 
Any order derivation $D$ is the sum of a
selfadjoint order derivation $D_1$ and a skew order derivation $D_2$;
with $a := D(\mathbb{I})$ choose $D_1 := R_a$ and $D_2 := D - D_1$.

This naming (selfadjoint and skew) stems from the fact that, in a von Neumann algebra, 
the selfadjoint order derivations have the shape $D(x) = (ax + xa)/2$ and 
the skew order derivations have the shape $D(x) = i(bx - xb)/2$, where $a,b$
are selfadjoint elements in the von Neumann algebra \cite{AS03}. 

The commutator $[D_1,D_2] := D_1 D_2 - D_2 D_1$
of any two order derivations $D_1$ and $D_2$ is an 
order derivation again and the order derivations 
thus form a Lie algebra \cite{AS01}. 
It is obvious that 
the commutator is skew if $D_1$ and $D_2$ are skew.
Therefore the skew order derivations form a Lie subalgebra $L$;
its elements are 
generators of one-parameter automorphism groups 
which describe reversible dynamical evolutions.
With any pair of elements
$a$ and $b$ in the order-unit space $A$, the operator 
$\left[R_a,R_b\right] - R_{b \Box a - a \Box b}$
lies in the Lie algebra $L$.

The next two lemmas provide some useful algebraic properties of the order automorphisms and
skew order derivations (similar to the situation in the more specific JB-algebras \cite{AS03}).
\\[0,3 cm]
Lemma 4: (i) If $W: A\rightarrow A$ is an order automorphism 
with $W(\mathbb{I})=\mathbb{I}$, then 
$W(a \Box b) = W(a) \Box W(b)$ for any $a,b \in A$.
\newline
(ii) If $D$ is a skew order derivation on $A$, then 
$D(a \Box b) = D(a) \Box b + a \Box D(b)$ for any $a,b \in A$.
\\[0,3 cm]
\textit{Proof.} (i) Suppose that $W$ is an order automorphism on $A$ 
with $W(\mathbb{I})=\mathbb{I}$. Then $W$ maps 
$ext\left[0,\mathbb{I}\right]$ onto itself.

For any state $\mu$ on $ext\left[0,\mathbb{I}\right]$,
$\mu W : e \rightarrow \mu(W(e))$ 
is a state on $ext\left[0,\mathbb{I}\right]$ and
$\mu(W(\cdot) | W(e))$ is a conditional probability under the event $e$ in the state $\mu W$. 
Its uniqueness (Lemma 2) implies that
$\mu(W(f) | W(e)) = \mu W (f | e)$ 
for any $e,f \in ext\left[0,\mathbb{I}\right]$ with $\mu(W(e)) > 0$.
That is $\mu(U_{W(e)} W(f)) = \mu(W(U_e f))$.
If $\mu(W(e)) = 0$, then 
$0 \leq \mu(U_{W(e)} W(f))\leq \mu(U_{W(e)} \mathbb{I}) = \mu(W(e)) = 0$
and $0 \leq \mu(W(U_e f)) \leq \mu(W(U_e \mathbb{I}) = \mu(W(e)) = 0$.
Thus $\mu(U_{W(e)} W(f)) = 0 = \mu(W(U_e f))$.

Therefore, 
$W(U_e f) = U_{W(e)} W(f)$ for any events $e$ and $f$; by Lemma 1 (iii),
it follows that $W (e \Box f) = W (e) \Box  W(f)$.
Since $A$ is the closed linear hull of the events, 
the continuity and linearity of the product and of $W$
finally imply $W (a \Box b) = W (a) \Box  W(b)$ for $a,b \in A$.

(ii) Suppose that $a,b \in A$ and that $D$ is a skew order derivation. By part (i)
$e^{tD}\left(a\Box b\right) = e^{tD}\left(a\right)\Box e^{tD}\left(b\right)$
for all real numbers $t$. Differentiating both sides of this equation at $t=0$
gives $D(a \Box b) = D(a) \Box b + a \Box D(b)$. $\blacksquare$
\\[0,3 cm]
Lemma 5: $\left[D,R_a\right] = R_{D (a)}$ for any skew order derivation $D$ and $a \in A$.
\\[0,3 cm]
Proof. Suppose $a,x \in A$. Then $D (x \Box a) = D (x) \Box a + x \Box D(a)$ by Lemma 4. 
This can be rewritten as $D R_a x  = R_a D x + R_{D (a)} x$. $\blacksquare$

\section {Dynamical correspondence}

In a von Neumann algebra $M$, there is the following one-to-one correspondence $a \rightarrow D_a$ between the 
selfadjoint elements $a$ of the algebra and the skew order derivations \cite{AS03}: 
$D_a x = i(ax - xa)/2$ for $x \in M$.
In this case, $\left[D_a,D_b\right] = - \left[R_a,R_b\right]$ and $D_a a = 0$ for all selfadjoint $a,b \in M$.
Moreover, this specific correspondence distinguishes those JB- and JBW-algebras that are the selfadjoint parts
of C*- and von Neumann algebras from the other ones. 
This motivates the following definition which is due to Alfsen and Shultz \cite{AS98, AS03},
but adapted to the more general setting of this paper.
Alfsen and Shultz consider only the JB- and JBW-algebras;
since these are commutative, they need not distinguish between 
the right-hand side and left-hand side multiplication operators
$R_a$ and $T_a$.

\textbf{Definition 1.} A \textit{dynamical correspondence} 
is a linear map $a \rightarrow D_a$ from $A$ 
into the Lie algebra $L$ of skew order derivations on $A$, which satisfies the 
following two conditions:
\newline
(i) $\left[D_a,D_b\right] = - \left[R_a,R_b\right]$ for $a,b \in A$,
\newline
(ii) $D_a a = 0$ for all $a \in A$.
\\[0,3 cm]
Condition (i) links the dynamical correspondence to the multiplication operation $\Box$
and immediately implies its commutativity. Applying
to $\mathbb{I}$ both sides of the equation gives $0 = b \Box a - a \Box b$.
Therefore, the operators $R_a$ and $T_a$ become identical.
Condition (i) thus has important mathematical consequences, 
but lacks any physical justification.
Condition (ii) means that $a$ is invariant under the one-parameter dynamical group 
generated by $D_a$.

It shall now be seen that the existence of a dynamical correspondence
implies not only the commutativity, but also the power-associativity 
and the Jordan property of the product. The following 
lemma and theorem and the proofs are transfers of the results 
in \cite{AS03} for JBW-algebras
to the more general setting of this paper.
Lemma 4 (ii) is the key making this possible.
\\[0,3 cm]
Lemma 6: Assume that the map $a \rightarrow D_a$ from $A$ into $L$
is a dynamical correspondence. Then 
\newline
(i) $\left[D_a,R_b\right] = \left[R_a, D_b\right]$ and 
\newline
(ii) $D_a b = - D_b a$ for any $a,b \in A$. 
\\[0,3 cm]
Proof. (i) Assume that the map $a \rightarrow D_a$ is a dynamical correspondence. 
By Lemma 5 and condition (ii) of Definition 1, 
$\left[D_a,R_a\right] = R_{D_a a} = R_0 = 0$ 
for all $a \in A$.
Therefore, by the linearity of the dynamical correspondence $a \rightarrow D_a$,
for all $a,b \in A$, $0 = \left[D_{a+b},R_{a+b}\right] = \left[D_a,R_b\right] + \left[D_b,R_a\right]$.
This gives $\left[D_a,R_b\right] = - \left[D_b,R_a\right] = \left[R_a,D_b\right]$.

(ii) Lemma 5 and (i) of Lemma 6 imply for all $a,b \in A$ that 
$D_a b = R_{D_a b} \mathbb{I} = \left[D_a, R_b\right]\mathbb{I} = \left[R_a, D_b\right]\mathbb{I} 
= - \left[D_b, R_a\right]\mathbb{I} = - R_{D_b a} \mathbb{I} = - D_b a$. $\blacksquare$
\\[0,3 cm]
\textbf{Theorem 1:} If $A$ admits a dynamical correspondence, it is (isomorphic to) the selfadjoint part
of an associative *-algebra over the complex numbers and the product $\Box$ becomes identical with the Jordan 
product: $a \Box b$ = $a \circ b$ = $(ab+ba)/2$ for $a,b \in A$
(This means that $A$ is a special Jordan algebra). 
\\[0,3 cm]
Proof. Assume that $a \rightarrow D_a$ is a dynamical correspondence on $A$. 
By Lemma 6, an anti-symmetric bilinear product $\times$ can be defined on $A$ via 
$a \times b := D_a b$ for $a,b \in A$. 
A further bilinear map into $A + iA$ (considered as a real-linear space) 
can be defined on $A$ via: $ab := a \Box b - i(a \times b)$.
This map can be uniquely extended to a bilinear product on $A + iA$ (considered as
a complex-linear space). It shall now be shown that this product is associative. Because of its
linearity, it suffices to prove that
$a(cb) = (ac)b$ for $a,b,c \in A$. This means
$$a \Box (c \Box b) - i(a \times (c \Box b)) - i(a \Box (c \times b)) - (a \times (c \times b))$$ 
$$= (a \Box b) \Box c - i((a \Box b) \times c) - i((a \times c) \Box b) - (a \times c) \times b.$$ 
Separating real and imaginary terms and using the anti-symmetry of the $\times$-product yields
the following two equations: 
$$a \times (b \times c) -  b \times (a \times c) = -a \Box (b \Box c) + b \Box (a \Box c)$$
and
$$ a \times (b \Box c) - b \Box (a \times c) = a \Box (b \times c) - b \times (a \Box c).$$
The left-hand side of the first one of these two equations is just $\left[D_a, D_b\right]c$
and its right-hand side is $-\left[R_a,R_b\right]c$. 
Note that the product $\Box$ is commutative and $T_x = R_x$ for $x\in A$,
which follows from the existence of the dynamical correspondence.
Similarly the left-hand side of the 
second equation is $\left[D_a,R_b\right]c$ and its right hand side is $\left[R_a,D_b\right]c$.
Thus the first one of these two equations follows directly from Definition 1 
and the second one from Lemma 6.

The involution on $A + iA$ is defined by $(a + ib)^{*} = a - ib$ and it must still
be shown that $(xy)^{*} = y^{*}x^{*}$ for $x,y \in A$. By linearity it suffices to prove that
$(ab)^{*} = ba$ for $a,b \in A$. This follows from the anti-symmetry of the $\times$-product,
since 
$$(ab)^{*} = \left(a \Box b - i\left(a \times b\right)\right)^{*} = a \Box b + i\left(a \times b\right) = b \Box a - i\left(b \times a\right) = ba.$$
Therefore, $A + iA$ is an associative *-algebra and its selfadjoint part is $A$. Moreover,
$(ab + ba)/2 = a \Box b$ for $a,b \in A$.
Note that the last equation again requires the commutativity of the product $\Box$. $\blacksquare$
\\[0,3 cm]
\textbf{Corollary 1:} If $A$ admits a dynamical correspondence 
and if $a^{2}\geq 0$ for any $a$ in $A$, 
then $A$ is (isomorphic to) the selfadjoint part of a C*-algebra.
\\[0,3 cm]
Proof. This follows immediately from Theorem 1 and 
Theorem 1.96 in Ref. \cite{AS01} (or A59 in Ref. \cite{AS03});
the C*-norm on $A + i A$ is given by $ \left\|x\right\| := \left\|x^{*}x\right\|^{1/2}$
for $x$ in $A + i A$. $\blacksquare$
\\[0,3 cm]
Theorem 1 shows that the assumption that a dynamical correspondence exists 
is very strong. From a general starting point, 
it immediately results in special Jordan algebras
which, moreover, are the selfadjoint parts of *-algebras over the complex numbers;
real algebras that cannot be obtained as selfadjoint parts of complex *-algebras are ruled out. 

However, Theorem 1 does not yet lead to ordinary Hilbert space quantum mechanics.
This is achieved by the additional
assumption that the squares of the elements in $A$ are positive. By Corollary 1,
$A$ is the selfadjoint part of a C*-algebra then and, 
by the Gelfand-Naimark Theorem \cite{AS01}, $A$ can be represented 
as operators on a complex Hilbert space.
In doing so, $A$ exhausts the full algebra of all operators 
on the Hilbert space in some cases or forms a genuine subalgebra
(as physically required with the presence of superselection rules) in other cases. 

Examples of algebras with positive squares, but without dynamical correspondences 
are the formally real Jordan algebras $H_n(\mathbb{R})$ and $H_n(\mathbb{H})$ ($n \geq 3$)
and the exceptional Jordan algebra $H_3(\mathbb{O})$. They consist
of the hermitian $n$$\times$$n$-matrices over the  
real numbers ($\mathbb{R}$), quaternions ($\mathbb{H}$) or octonions ($\mathbb{O}$) \cite{AS03}.
Examples with dynamical correspondences, but with non-positive squares are not known.

\section {Conclusions}

In order to figure out why quantum physics needs the complex Hilbert space, 
many attempts have been made to distinguish the C*-algebras and von Neumann algebras from 
the more general JB- and JBW-algebras. Different distinguishing
properties have been identified: dynamical correspondences, the 3-ball property and orientations \cite{AS01, AS03}.
Only the dynamical correspondence has a certain physical meaning, since it establishes a relation
between the algebra elements and the bounded generators of one-parameter dynamical groups.
However, only the existence of (possibly unbounded) group generators can be considered an important requirement
for any reasonable physical theory, and theories without dynamical correspondences or with less strong versions
might be thinkable.
Alfsen and Shultz's conditon (i) in Definition 1 (section 5) links the dynamical correspondence 
to the multiplication operation $\Box$. It is a mathematically nice and strong
assumption, but lacks a proper physical justification. The \textit{energy observable assignment}
defined in \cite{Bar14} represents a weaker form of a dynamical correspondence
dispensing with condition (i). It may be better justified from the physical point of view,
but the mathematical methods applied in \cite{Bar14} fail in the infinite-dimensional case
and it is not known whether the results in \cite{Bar14} remain valid in this case.

In the present paper, Alfsen and Shultz's definition of a dynamical correspondence has been used and 
it has been seen that this notion
can be extended to a class of nonassociative algebras, which is much broader than the JBW algebras.
This class arises from recent studies of the quantum logics with a conditional probability calculus 
(i.e., with a reasonable model of the L\"uders - von Neumann quantum measurement process)
and particularly of those that rule out third-order interference. The existence of 
a dynamical correspondence for an algebra in this class still entails that it is the selfadjoint
part of a C*-algebra, if it is assumed that the squares of the algebra elements are positive (Corollary 1). 
The Jordan property of the product or its power-associativity 
become redundant requirements in this situation. The same holds for some other 
conditions used for abstract mathematical characterizations of operator algebras
or their state spaces (e.g., spectrality and ellipticity \cite{AS03}).

Thus, within the considered class of nonassociative algebras, the two postulates that a dynamical
correspondence exists and that the square of any algebra element is positive
result in ordinary quantum mechanics 
using the complex Hilbert space or, vice versa, a genuine generalization of 
quantum theory must omit at least one of them.

In section 4, it has been seen that the skew order derivations form a Lie algebra.
Almost all finite-dimensional simple Lie algebras arise 
from the derivations on the finite-dimensional formally real Jordan algebras
(the finite-dimensional version of the JB-/JBW-algebras),
and there are only four exceptions 
($\mathfrak{g}_2$, $\mathfrak{e}_6$, $\mathfrak{e}_7$ and $\mathfrak{e}_8$ \cite{Baez01}).
An interesting question now becomes
whether these four emerge from the skew order derivations on some unknown nonassociative algebras 
out of the class which is defined in the second section and comprises
the finite-dimensional formally real Jordan algebras. 
If such a nonassociative algebra exists, 
it either contains elements with non-positive squares
or does not possess a dynamical correspondence, and its
continuous symmetries form one of the exceptional Lie groups $G_2$, $E_6$, $E_7$ and $E_8$.

\end{document}